\documentclass[12pt,a4paper]{article}
\usepackage[T1]{fontenc}
\usepackage[utf8]{inputenc}
\usepackage{authblk}

\title{\vspace{2.2cm}\begin{flushleft} \bf {{Quantum Entropy for the Fuzzy Sphere and its Monopoles} 
\linethickness{.05cm}\line(1,0){433}
}\end{flushleft}} 

\author[a]{\bf Nirmalendu~Acharyya \footnote{nirmalendu@cts.iisc.ernet.in}}
\author[b]{\bf Nitin~Chandra \footnote{nitinc@imsc.res.in}}
\author[a]{\bf Sachindeo~Vaidya\footnote{vaidya@cts.iisc.ernet.in}}
\affil[a]{ \small Centre for High Energy Physics,  Indian Institute of Science, Bangalore-560012,
India}
\affil[b]{ \hspace{-.2cm} The Institute of Mathematical Sciences, C.I.T. Campus, Taramani, Chennai-600113, India}

\usepackage{amsmath,amssymb,epsfig,cite,dsfont}
\usepackage{array,multirow}
\usepackage {color}
\usepackage[toc,page]{appendix}
\numberwithin{equation}{section}

\usepackage[colorlinks=true
,urlcolor=blue
,anchorcolor=blue
,citecolor=blue
,filecolor=blue
,linkcolor=blue
,menucolor=blue
,pagecolor=blue
,linktocpage=true
,pdfproducer=medialab
,pdfa=true
]{hyperref}

\advance\hoffset by -1.1truecm
\advance\voffset by -1.0truecm
\newcommand{\be}{\begin{equation}}
\newcommand{\ee}{\end{equation}}
\newcommand{\bea}{\begin{eqnarray}}
\newcommand{\eea}{\end{eqnarray}}

\date{}
\newcommand\bigzero{\makebox(0,0){\text{\huge0}}}
\newcommand*{\abord}{\multicolumn{1}{|c}{}}
\newcommand*{\bbord}{\multicolumn{1}{|c}}

\usepackage[all]{xy}

\numberwithin{equation}{section}
\def\boxed#1{
  \leavevmode\thinspace
  \hbox{\vrule\vtop{\vbox{\hrule\kern1pt
        \hbox{\thinspace$\displaystyle{#1}$\thinspace}}
      \kern1pt\hrule}\vrule}\thinspace}
\newcounter{appendice}

\begin{document}
\maketitle

\abstract{Using generalized bosons, we construct the fuzzy sphere $S_F^2$ and monopoles on $S_F^2$ in a reducible representation of $SU(2)$. The corresponding quantum states are naturally obtained using the GNS-construction. We show that there is an emergent non-abelian unitary gauge symmetry which is in the commutant of the algebra of observables. The quantum states are necessarily mixed and have non-vanishing von Neumann entropy,  which increases monotonically under a bistochastic Markov map. The maximum value of the entropy has a simple relation to the degeneracy of the irreps that constitute the reducible representation that underlies the fuzzy sphere.} 

\newpage
 \tableofcontents

\section{Introduction}

Fuzzy spaces are some of the simplest noncommutative geometries. These emerge from the observation that coadjoint orbits of Lie groups are symplectic manifolds and can be quantized under certain conditions. For such quantized spaces, precise localization of a point is not possible and thus these spaces are fuzzy. Mathematically, the fuzzy space is described by the algebra of the linear operators on the representation spaces of the Lie group. As such operator algebras are noncommutative, the algebra of ``functions'' on the space is noncommutative. Further, if the classical manifold which is the orbit of the Lie group is compact, the representations of the group are finite dimensional. In this case the fuzzy space is a finite dimensional matrix algebra on which the group acts in a simple way.  The continuum manifold can be obtained as a classical limit by taking the ``effective Planck's constant'' of the quantization to zero.

For example, the 2-sphere is a orbit of $SU(2)$ through the pauli matrix $\sigma_3$ and has a symplectic form. $S^2$ is the set of points 
\begin{equation}\nonumber
g\sigma_3 g^{-1}, \quad\quad  g \in SU(2).
\end{equation}
The 2-sphere can be described by two angles $\theta, \phi: 0\leq\theta\leq \pi,$ $0\leq\phi<2 \pi$, while the symplectic form is 
$$\mathcal{J} d \cos\theta \wedge d\phi.$$
When quantized, the fuzzy sphere $S^2_F$ is described by the matrix algebra on the spin-$\mathcal{J}$ representation of the $SU(2)$. As the spin-$\mathcal{J}$ representation space of the $SU(2)$ is $(2\mathcal{J}+1)$-dimensional, $S^2_F$ is described by the algebra of $(2\mathcal{J}+1)\times (2\mathcal{J}+1)$ square matrices.
The elements of this algebra are the identity $\mathbb{I}_{(2\mathcal{J}+1)}$, the generators of $SU(2)$ $\hat{x}_i$ $(i=1,2,3)$ and products of these generators. More precisely, the matrix algebra 
\begin{equation}\label{fuzzy_algebra_213}
 [\hat{x}_i,\hat{x}_j]=i\epsilon_{ijk}\hat{x}_k, \quad \hat{x}_i^\dagger =\hat{x}_i, \quad \hat{x}_i\hat{x}_i=\mathcal{J}(\mathcal{J}+1)\mathbb{I}_{(2\mathcal{J}+1)}, \quad i,j,k=1,2,3.
\end{equation}
is sufficient describe the fuzzy sphere $S_F^2$.

The continuum 2-sphere is obtained in the limit $\mathcal{J} \rightarrow \infty$. In order that the algebra of $S_F^2$ have the correct continuum limit, it is more appropriate to work instead with $\hat{s}_i=\frac{1}{\mathcal{J}} \hat{x}_i$  \cite{{Hoppe:1982},{Madore:1991bw}} obeying
\begin{equation}\label{fuzzy_algebra_214}
 [\hat{s}_i,\hat{s}_j]= \frac{i}{\mathcal{J}} \epsilon_{ijk} \hat{s}_k, \quad \hat{s}_i^\dagger =\hat{s}_i, \quad \hat{s}_i\hat{s}_i=(1+ \frac{1}{\mathcal{J}})\mathbb{I}_{(2\mathcal{J}+1)}, \quad i,j,k=1,2,3.
\end{equation}
It is easy to see that (\ref{fuzzy_algebra_214}) in this limit gives the ordinary or commutative 2-sphere.

Such  discrete spaces retain the symmetries of the underlying classical manifold and hence are interesting from the point of view of mathematics 
and physics.
Field theories on such a compact fuzzy space are finite dimensional and do not require UV regularization.  Additionally, these theories can naturally incorporate topological objects like instantons and axial anomalies.
Hence the nontrivial field configurations on such spaces, especially  the classical topological objects like solitons, instantons and monopoles have been the subject of paricular interest to many (for example see\cite{{Grosse:1995jt},{Grosse:1995pr},{Grosse:1996mz},{Douglas:1997fm},{Baez:1998he},   {Balachandran:1999hx}, {Vaidya:2001rf},{Acharyya:2013hga}}).

These spaces emerge naturally in the matrix models describing branes.  For example, in the M2-M5 brane system the transverse geometry is a fuzzy space (see \cite{Basu:2004ed,Nastase:2009ny}). Fuzzy spaces also appear in the dynamics of the D-branes in non-trivial backgrounds (see for instance \cite{{Alekseev:2000fd},{Hashimoto:2001xy}, {Tomino:2003hb}} and references therein).

An interesting model of this type was first discussed in  \cite{Myers:1999ps}. Roughly speaking, it describes a three-matrix model coupled to a background Ramond-Ramond 4-form field and is described by the action
\begin{eqnarray}\label{matrix_action}
S = T_0 Tr\left[\frac{1}{2}\dot{ \phi_i}^2+\frac{1}{4} [\phi_i, \phi_j]^2 -\frac{i}{3}\kappa \epsilon_{ijk} \phi_i[\phi_j, \phi_k]\right].
\end{eqnarray}
$\kappa$ is a (Chern-Simons) coupling constant and $\phi_i$'s ($i=1,2,3$) are $N\times N$ matrices. 
This model has been used widely in understanding the physics of branes. For example in \cite{Polchinski:2000uf}, this is used to study the super-Yang-Mills theory in four dimensions. The importance of classical nonpertubative solutions like flux tubes, instantons etc in these matrix models are also emphasized  in \cite{Polchinski:2000uf}.

In absence of the Chern-Simons term (i.e. when $\kappa=0$), the potential is extremized by
\begin{equation}
[\phi_i, \phi_j]=0. 
\end{equation}
This extremum represents $N$ $D0$-branes.
But in presence of the Chern-Simons term, there are other extrema of lower energy. These extrema are given by a set of noncommuting matrices
\begin{eqnarray}
\phi_i = \kappa\hat{x}_i, \quad \quad [\hat{x}_i,\hat{x}_j] = i \epsilon_{ijk} \hat{x}_k, \quad\textrm{for } i,j,k=1,2,3.
\end{eqnarray}
This noncommutative solution is interpreted as $N$ $D0$-branes attached to a spherical $D2$-brane. This configuration has zero $D2$-charge but there is a possible nonzero finite dipole coupling. 

Here,  $\hat{x}_i$'s are $N\times N$ matrices satisfying $\hat{x}_i \hat{x}_i =\textrm{fixed}$ and thus describes  the algebra of $S_F^2$ on a spin-$\mathcal{J} (\equiv \frac{N-1}{2})$ representation of $SU(2)$. For the model (\ref{matrix_action}), this fuzzy sphere solution can either be an irreducible or reducible representation.  If the representation is reducible (i.e. $N\times N$ matrices are block diagonal with the blocks of smaller size), the classical energy $E_{nc}^{r}$ is higher than that of the irreducible one $E_{nc}^{ir}$. 


A careful analysis of  stability done around the critical points of the energy function in {\cite{Jatkar:2001uh}} shows that all these extremas (both the reducible and the irreducible) are stable. This is a puzzling circumstance as the fuzzy spheres in the reducible representation have higher energy. In {\cite{Jatkar:2001uh,Kimura:2003kd}},  the puzzle is resolved by taking non-spherical marginal deformations around the fuzzy spheres in reducible representation. The system now have tachyonic instabilites and the cascading described in  \cite{Myers:1999ps} is a roll down by tachyon condensations.

In this article, we investigate this problem from a completely different perspective. We will show that when the irreps in the reducible representation are identical, all fuzzy spheres in  various irreps have the same ``radius" (see table 1 in {\cite{Jatkar:2001uh}). Thus it is possible to identify the reducible algebra with a single fuzzy sphere in a reducible representation.  We construct  these fuzzy spheres  by Schwinger construction.  To do so, we take the 2-dimensional oscillators in reducible representations.  Such oscillators are known as Brandt-Greenberg \cite{bg} or generalized Bose oscillators. In \cite{Acharyya:2011bx}, it was shown that these play a vital role in the construction of classical topological solutions in noncommutative spaces. Here we show that these oscillators give the monopoles on the background of the fuzzy sphere with reducible representation. A remarkable consequence is that the quantum state corresponding to these monopoles is not pure, and hence carries intrinsic quantum entropy \cite{{Balachandran:2012pa},{Balachandran:2013kia}}. This entropy persists at zero temperature, and can be macroscopically large.

The article is organized as follows. In section \ref{section_1}, we briefly review the description of fuzzy sphere algebra using the standard Schwinger construction and the technique developed in \cite{Grosse:1995jt} to describe the sections of the complex line bundle on the fuzzy sphere. In section \ref{section_2} we review the Brandt-Greenberg oscillators in detail and in section \ref{section_3} the Schwinger construction with these oscillators. We show that the reducible representation thus obtained describe a fuzzy 2-sphere. 
In section \ref{section_4} we construct the associated line bundles on the $S_F^2$ corresponding to these unstable vacua using the prescription in \cite{Grosse:1995jt}.
In section \ref{section_6}, we construct the quantum states of the fuzzy sphere in the reducible representation. We show that these have a $U(r)$ gauge symmetry ($r$ is the number of identical irreps in the reducible representation) and are impure.  Similarly, the states of the monopole have a $U(r^2)$ gauge symmetry and are impure as well. We compute the entropy associated with these impure states.

\section{A brief review - Fuzzy sphere and associated line bundle}\label{section_1}

%

There is a natural chain of descent $\mathbb{C}_F^2 \rightarrow S^3_F \rightarrow S^2_F$. 
The algebra of $\mathbb{C}_F^2$ is described by the algebra of a pair of independent harmonic oscillators (see \cite{Balachandran:2005ew})
\begin{eqnarray}
[\hat{a}_\alpha , \hat{a}_\beta^\dagger] = \delta_{\alpha \beta}, \quad \quad \alpha, \beta= 1,2.
\end{eqnarray}
These oscillators acts on the Fock space $\mathcal{F}= span\{| n_1, n_2\rangle \}$. These are eigenstates of $\hat{N}= \sum_\alpha \hat{a}_\alpha^\dagger \hat{a}_\alpha$.

The operator $\hat{\chi}_\alpha=\hat{ a}_\alpha\frac{1}{\sqrt{\hat{N}}}$ is well defined on $\mathcal{F}$ except for the state $|0,0 \rangle$. As $\hat{\chi}_\alpha^\dagger \hat{\chi}_\alpha=1$, the algebra generated by $\hat{\chi}_\alpha$  describes the  fuzzy 3-sphere $S^3_F$. 

The Schwinger construction 
\begin{equation}\label{sch_1}
\hat{x}_i = \frac{1}{2}\hat{a}_\alpha^\dagger (\sigma_i)_{\alpha\beta} \hat{a}_\beta, \quad [\hat{x}_i, \hat{x}_j] = i \epsilon_{ijk} \hat{x}_k, \quad \hat{x}_i \hat{x}_i = \frac{\hat{N}}{2} \left(\frac{\hat{N}}{2}+1\right). 
\end{equation}
is the noncommutative version of Hopf map \cite{Grosse:1992bm}. 

In the subspace $\mathcal{F}_n=span\{| n_1, n_2\rangle: \, n_1 +n_2=n \} \subset \mathcal{F}$, the operator $\hat{x}_i \hat{x}_i = \frac{n}{2} \left(\frac{n}{2}+1\right)$= fixed. Thus this subspace is the carrier space of $(n+1)$ dimensional UIR of $SU(2)$.

The operators
\begin{equation}
\hat{q}_i = \frac{1}{\hat{N}} \hat{x}_i
\end{equation}
generate the algebra of the fuzzy sphere $S_F^2$.
The map 
\begin{equation}
\hat{q}_i = \frac{1}{2}\hat{\chi}_\alpha^\dagger  (\sigma_i)_{\alpha\beta} \hat{\chi}_\beta
\end{equation}
 from $S_F^3 \rightarrow S_F^2$ is the noncommutative analogue of the Hopf fibration.

The associated complex line bundles are given by complex scalar fields $\Phi$ with a topological charge. These scalar fields  map $\mathcal{F}_n \rightarrow \mathcal{F}_l$, where the topological charge is $\kappa= \frac{1}{2}(l-n)$. 
We can find a basis for these $\Phi$ exploiting the group theoretic properties of $S_F^2$ \cite{Grosse:1995jt}.


In general, $\Phi$'s are $(l+1)\times (n+1)$ rectangular matrices and are element of a noncommutative bi-module $\mathcal{H}_{nl}$ -- it is a left $\mathcal{A}_l$-module and a right $\mathcal{A}_n$-module.
On $\mathcal{H}_{nl}$, the adjoint of $\hat{x}_i$ acts as
\begin{eqnarray}\label{su2_action_12}
Ad(\hat{x}_i)\Phi = \left[\hat{x}_i^{(l)} \Phi - \Phi \hat{x}_i^{(n)}\right], \quad \quad \Phi \in \mathcal{H}_{nl},
\end{eqnarray}
and generates rotation
\begin{equation}
\Big[Ad(\hat{x}_i),Ad(\hat{x}_j)\Big] = i\epsilon_{ijk} Ad(\hat{x}_k).
\end{equation}
 This action of the $SU(2)$ gives the direct product  $\frac{l}{2}\otimes \frac{n}{2}$ of the  two UIRs $\frac{l}{2}$ and $\frac{n}{2}$. The elements of  $\mathcal{H}_{nl}$ can therefore be expanded in terms of the eigenfunctions $\Psi^j_{J,\kappa,m}$ of $Ad(\hat{x}_3)$ and $[Ad(\hat{x}_i)Ad(\hat{x}_i)]$ belonging to the irreducible representations in the decomposition of $\frac{l}{2}\otimes \frac{n}{2}$:
\begin{equation}
\frac{l}{2}\otimes \frac{n}{2} =\kappa\oplus (\kappa+1)\oplus\ldots\oplus J, \quad \quad \textrm{where } J \equiv \frac{ l+n}{2}.
\end{equation}

%
%
%
%
%
%
%
%
%
An arbitrary element $\Phi$ of $\mathcal{H}_{nl}$ can be expressed as
\begin{equation}
\Phi= \sum_{j=\kappa}^J \sum_{m=-j}^j c^j_{J,\kappa,m}\Psi^j_{J,\kappa,m}, \quad c^j_{J,\kappa,m} \in \mathbb{C}.
\end{equation}

The topological charge operator $\hat{K}_0$ is
\begin{equation}
\hat{K}_0 \equiv \frac{1}{2} \left[\hat{N}, \,\,\, \right]
\end{equation}
and it satisfies
\begin{eqnarray}
&\left[ Ad(\hat{x}_3) , \hat{K}_0 \right] =0= \left[ Ad(\hat{x}_i)Ad(\hat{x}_i) , \hat{K}_0 \right].
\end{eqnarray}
Any element $\Phi$ of $\mathcal{H}_{nl}$ is also an eigenfunction of  $\hat{K}_0$:
\begin{equation}
\hat{K}_0\Phi= \frac{\kappa}{2} \Phi.
\end{equation}
$\Phi$ is thus the noncommutative analogue of a section of the complex line bundle with topological charge $\kappa$.

\section{Generalized Bosonic Oscillators}\label{section_2}
The generalized bosonic oscillators \cite{bg}  change the number of quanta by a positive integer $K$. We briefly recall their construction in this section.

Consider a standard bosonic oscillator described by the  operators $\hat{a}^\dagger$ and $\hat{a}$ satisfying
\begin{equation}
[\hat{a},\hat{a}^\dagger]=1.
\label{basiccomm}
\end{equation} 
The  Hilbert space $\mathds{H}$ is spanned by the 
 basis of the eigenvectors of the {\it number} operator $\hat{N} \equiv \hat{a}^\dagger \hat{a}$: 
$$\mathds{H}=\{c_n|n \rangle, n=0,1,\cdots, \infty \}, \quad\quad \hat{N}|n\rangle = n |n\rangle,$$
and $\hat{a}$ and $\hat{a}^\dagger$ act on this basis in the standard way.

Thus $(\hat{a},\mathds{H})$ is a representation of the oscillator 
algebra (\ref{basiccomm}). It is also the unique (up to unitary equivalence) irreducible representation of this 
algebra (for example see \cite{putnam}).

$\mathds{H}$ can be split into two disjoint subspaces 
$\mathds{H}_e=\{\sum c_{2n}|2n\rangle  \in \mathds{H}\}$ and $\mathds{H}_o 
=\{\sum c_{2n+1}|2n+1\rangle  \in \mathds{H}\}$ . (The labels $e$ and $o$ stand for \textit{even} and \textit{odd} respectively.) The (projection) operators
\begin{equation}\label{proje_1}
\Lambda^e = \sum_{n=0}^\infty |2n\rangle\langle2n|, \quad \Lambda^o = \sum_{n=0}^\infty |2n+1 \rangle \langle 2n+1|
\end{equation}
project onto the subspaces $\mathds{H}_e$ and $\mathds{H}_o$ respectively. 
On  $\mathds{H}_{e}$,  one can define  operators $\hat{b}^e$ 
and   $\hat{b}^{e\dagger}$ as 
\begin{eqnarray}
 \hat{b}^e |2n\rangle = n^{\frac{1}{2}} |2n-2\rangle, \quad   \hat{b}^{e\dagger} |2n\rangle = (n+1)^{\frac{1}{2}} |2n+2\rangle,   \quad
\hat{b}^e |0\rangle =0.
\end{eqnarray}
 Similarly on $\mathds{H}_{o}$, we can define $\hat{b}^{o}$ and $\hat{b}^{o\dagger}$ as 
\begin{eqnarray}
\hat{b}^o|2n+1\rangle = n^{\frac{1}{2}} |2n-1\rangle,  \quad \hat{b}^{o\dagger}|2n+1\rangle = (n+1)^{\frac{1}{2}} |2n+3\rangle, \quad \hat{b}^o |1\rangle =0.
\end{eqnarray}
These operators satisfy $[\hat{b}^e, \hat{b}^{e\dagger} ]=1$ and $[\hat{b}^o, \hat{b}^{o\dagger} ]=1$.
Thus as representations $(\hat{b}^o,\mathds{H}_o)$, $(\hat{b}^e,\mathds{H}_e)$ and $(\hat{a},\mathds{H})$ are isomorphic to 
each other. In other words, there exist unitary operators $U_e$ and $U_o$ such that $U_e^\dagger  \hat{b}^e
U_e= \hat{a}$ and $U_o^\dagger  \hat{b}^o
U_o= \hat{a}$.

Note that all these operators are unbounded and  are hence defined on a dense domains in $\mathds{H}$. A detailed discussion about these domains can be found in \cite{bg}.

Using $\Lambda^e$ and $\Lambda^o$ of (\ref{proje_1}) , one can define an operator $\hat{b}$ 
\begin{equation}
 \hat{b}=\hat{b}^e\Lambda^e +\hat{b}^o\Lambda^o
\label{irreducible_decomp}
\end{equation}
whose action  on the basis vectors $|n\rangle$ is
\begin{equation}
 \hat{b}|2n\rangle = n^\frac{1}{2} |2n-2\rangle, \quad \hat{b}|2n+1\rangle = n^\frac{1}{2} |2n-1\rangle
\label{action_b}.
\end{equation}
Notice that both $|0\rangle$ and $|1\rangle$ are annihilated by $\hat{b}$.

The operator $\hat{b}$ satisfies the commutation relation $[\hat{N}, \hat{b}] =-2\hat{b}$. A new number operator  $\hat{M}\equiv\hat{b}^\dagger \hat{b}= \frac{1}{2}\left(\hat{N}-\Lambda^o\right)$  has  $\left\{|n\rangle\right\}$  as eigenstates but each eigenvalue 
is two-fold degenerate. We can denote these eigenvalues  by $m_n=\frac{1}{2} (n-\lambda) $ where $\lambda$ takes values $0$ and $1$ for even and odd $n$'s respectively. Then (\ref{action_b}) can be written as
\begin{equation} \label{action_b_1}
 \hat{b}|n\rangle=m_n^\frac{1}{2}|n-2\rangle \quad \mathrm{and} \quad  \hat{b}^\dagger|n\rangle=(m_n+1)^\frac{1}{2}|n+2\rangle.
\end{equation}
The operator $\hat{b}$  satisfies $[\hat{b},\hat{b}^\dagger] =1$ and thus ($\hat{b},
\mathds{H}$) forms a reducible representation of the oscillator algebra, with (\ref{irreducible_decomp}) as its
decomposition into  irreducibles.

The above can be generalized to construct an operator $\hat{b}^{(K)}$ which lowers a state 
$|n\rangle$ by $K$ steps. We start by defining projection operators $\Lambda^i$ by
\begin{equation}
\Lambda^i = \sum_{n=0}^\infty |K n +i\rangle \langle Kn + i |, \quad i = 0,1, \cdots K-1.
\end{equation} 
that project onto subspaces $\mathds{H}_i =\{\sum_n c_{Kn+i} |K n +i\rangle \in \mathds{H} \}$.
In each  $\mathds{H}_i$, we define operators $\hat{b}^{i}$ and 
$\hat{b}^{i\dagger}$ that satisfy $[\hat{b}^{i}, \hat{b}^{i\dagger}]=1$ and hence correspond to the UIR of the oscillator algebra. A reducible representation is given by
\begin{equation}
 \hat{b}^{(K)} = \sum_{i=0}^{K-1} \hat{b}^i \Lambda^i, \quad \hat{b}^i |Kn+i\rangle = \sqrt{n} |Kn+i-K\rangle, 
 \quad \mathds{H}=\oplus_{i=0}^{K-1}\mathds{H}_i
\label{b_red}
\end{equation}
with $[\hat{b}^{(K)},\hat{b}^{(K)\dagger}]=1$. Again,$(\hat{b}^i,\mathds{H}_i)$ is isomorphic to $(\hat{a},\mathds{H})$ and $(\hat{b}^{(K)},\mathds{H})$ 
forms a reducible representation of the oscillator algebra.

The equations (\ref{irreducible_decomp})--(\ref{action_b_1}) represent the case $K=2$, the simplest non-trivial example of this construction.
Henceforth we will use $\hat{b}$ for $\hat{b}^{(2)}$. An explicit expression for $\hat{b}$ is \cite{katriel} 
\begin{equation}
 \hat{b}= \frac{1}{\sqrt{2}} \left( \hat{a}\frac{1}{\sqrt{\hat{N}}}\hat{a}\Lambda^e +\hat{a}\frac{1}{\sqrt{\hat{N}+1}}\hat{a}\Lambda^o\right)
\label{expression_b}
\end{equation}

The discussion here is telegraphic,  details may be found in \cite{bg, katriel}. 

\section{Schwinger construction with Generalized Bose Operators}\label{section_3}

By using the $\hat{b}^{K}$'s of the previous section in the Schwinger construction, we  get reducible representations of $SU(2)$.  This will have non-trivial  implications in the construction of the line bundles on the fuzzy sphere. Let us briefly see this.

The fuzzy space $\mathbb{C}^2_F$  is described by  two independent oscillators
which acts on the space $\mathcal{F}$
\begin{eqnarray}
\mathcal{F} = \mathds{H} \otimes \mathds{H}
\end{eqnarray}
This space can be spanned by the eigenstates of the \textit{number operator} $\hat{N}=\sum_\alpha \hat{a}_\alpha^\dagger \hat{a}_\alpha$:
\begin{equation}
\mathcal{F}=span\{|n_1, n_2\rangle: \hat{N}_\alpha |n_1, n_2\rangle =  n_\alpha |n_1, n_2\rangle,\,\, \alpha=1,2\}, \quad \quad
\end{equation}
where
\begin{equation}
 |n_1, n_2\rangle \equiv |n_1\rangle \otimes |n_2\rangle.
\end{equation}

$\mathcal{F}$ can now be split into four subspaces 
\begin{eqnarray} \left.\begin{array}{cclll}
\mathcal{F}^{ee}&=&\mathds{H}_e \otimes \mathds{H}_e=span\{|2n_1, 2n_2\rangle\} \\ \\
\mathcal{F}^{eo}&=&\mathds{H}_e \otimes \mathds{H}_o=span\{|2n_1, 2n_2+1\rangle\} \\ \\
\mathcal{F}^{oe}&=&\mathds{H}_o \otimes \mathds{H}_e=span\{|2n_1+1, 2n_2\rangle\} \\\\
\mathcal{F}^{oo}&=&\mathds{H}_o \otimes \mathds{H}_o=span\{|2n_1+1, 2n_2+1\rangle\} 
\end{array}\right\} \quad \textrm{ i.e. }\mathcal{F}= \oplus_{\lambda}\mathcal{F}^\lambda
\end{eqnarray}
where $\lambda=ee, eo, oe, oo$.

The projectors 
\begin{eqnarray}
&\Lambda_{ee} = \Lambda_{1}^e \Lambda_{2}^{e}, \quad \Lambda_{eo} = \Lambda_{1}^{e} \Lambda_{2}^{o}, \quad \Lambda_{oe} = \Lambda_{1}^{o} \Lambda_{2}^{e}, 
\quad \Lambda_{oo} = \Lambda_{1}^{o} \Lambda_{2}^{o}, 
\end{eqnarray}
\begin{eqnarray}
&\sum_{\lambda}  \Lambda_{\lambda} = 1, \quad\quad  \Lambda_\lambda \Lambda_{\lambda^\prime}= \Lambda_{\lambda} \delta_{\lambda\lambda^\prime}.
\end{eqnarray}
 project to subspaces of the two oscillator Hilbert space: $\Lambda_{\lambda}$ projects to 
subspace $\mathcal{F}^{\lambda}$.
($\Lambda_\alpha^e$ and $\Lambda_\alpha^o$ are the projectors  defined in (\ref{proje_1}).)
Explicitly 
\begin{equation}
\Lambda_{ee} = \cos^2 \frac{\hat{N}_1 \pi}{2} \cos^2 \frac{\hat{N}_2 \pi}{2}, \quad \Lambda_{eo} = \cos^2 \frac{\hat{N}_1 \pi}{2} \sin^2 \frac{\hat{N}_2 \pi}{2} \quad etc. 
\end{equation}

Consider two independent generalized bosonic oscillator $\hat{b}_1$ and $\hat{b}_2$: 
\begin{equation}
\hat{b}_\alpha = \hat{b}^e_\alpha \Lambda_\alpha^{e}+\hat{b}^o_\alpha \Lambda_\alpha^{o}, \quad\quad \alpha=1,2
\end{equation}
where
\begin{equation}
\hat{b}^e_\alpha= \frac{1}{\sqrt{2}} \hat{a}_\alpha \frac{1}{\sqrt{\hat{N_\alpha}}}\hat{a}_\alpha, \quad \hat{b}^o_\alpha= \frac{1}{\sqrt{2}} \hat{a}_\alpha \frac{1}{\sqrt{\hat{N}_\alpha+1}}\hat{a}_\alpha, \quad\quad \hat{N}_\alpha = \hat{a}_\alpha^\dagger \hat{a}_\alpha.
\end{equation}
$\hat{b}^e_1$ only acts on the states in $\mathcal{F}^{ee} \oplus \mathcal{F}^{eo}$ and so on.
Let us define a set of operators
\begin{equation}
\hat{\xi}_\alpha =\hat{b}_\alpha \frac{1}{\sqrt{\hat{M}}} , \quad \quad \alpha=1,2, \quad \quad \hat{M}=\hat{b}_1^\dagger \hat{b}_1+\hat{b}_2^\dagger \hat{b}_2.
\end{equation}
The operator $\hat{M}$
\begin{equation}
\hat{M}= \frac{1}{2}\left(\hat{N} - \Lambda_1^o - \Lambda_2^o\right)
\end{equation}
 vanishes when acted on the states $|0,0\rangle$, $|0,1\rangle$, $|1,0\rangle$ and $|1,1\rangle$. We exclude these states from the domain of $\hat{\xi}_\alpha$ so that $\hat{\xi}_\alpha$ are well-defined. These operators satisfy 
\begin{equation}
\hat{\xi}_\alpha^\dagger \hat{\xi}_\alpha = 1
\end{equation}
and hence defines the fuzzy 3-sphere $S_F^3$.

We define the operator map 
\begin{equation}\label{js_n}
\hat{t}_i = \frac{1}{2} \hat{\xi}_\alpha^\dagger \left(\sigma_i\right)_{\alpha\beta} \hat{\xi}_\beta= \frac{1}{\hat{M}} \hat{s}_i, \quad\quad \hat{s}_i = \frac{1}{2} \hat{b}^\dagger \sigma_i \hat{b},\quad\quad i=1,2,3
\end{equation}
where
\begin{eqnarray}
\hat{b} = \left(\begin{array}{lll}
\hat{b}_1\\
\hat{b}_2
\end{array}\right) \quad \textrm{and} \quad \sigma_i = \textrm{ Pauli matrices}.
\end{eqnarray}

They satisfy
 \begin{eqnarray}\label{nc_fuzzy}
\left[\hat{s}_i, \hat{s}_j \right] = i \epsilon_{ijk} \hat{s}_k, \quad \left[\hat{s}_k, \hat{N}\right]=0= \left[\hat{s}_k, \hat{M}\right].
\end{eqnarray}

The ``Casimir'' can be expressed as
\begin{equation}
\hat{s}_i \hat{s}_i = \frac{\hat{M}}{2} \left(\frac{\hat{M}}{2}+1\right), \quad \quad \hat{t}_i \hat{t}_i = \frac{1}{2} \left(\frac{1}{2} + \frac{1}{\hat{M}}\right).
\end{equation}

Let us define the space $\mathcal{G}_n$
\begin{eqnarray}\label{split_g_n}
\mathcal{G}_n = \mathcal{F}_{2n}^{ee} \oplus \mathcal{F}_{2n+1}^{eo}\oplus \mathcal{F}_{2n+1}^{oe}\oplus \mathcal{F}_{2n+2}^{oo}, 
\quad\quad \mathcal{G}_n \subset \mathcal{F}. \label{g_n_space}
\end{eqnarray}
In this subspace,  $\hat{M}$ takes the value $n$  and hence in $\mathcal{G}_n$, $\hat{t}_i\hat{t}_i$ is fixed :
\begin{eqnarray}
 \hat{t}_i \hat{t}_i = \frac{1}{2} \left(\frac{1}{2} + \frac{1}{n}\right).
\end{eqnarray}
The algebra generated by $\hat{t}_i$ restricted to $\mathcal{G}_n$  is the fuzzy 2-sphere $S^2_F$ and the Jordan-Schwinger construction (\ref{js_n}) is the fuzzy analogue of the the Hopf map: $S^3_F \rightarrow S^2_F$.

In the commutative limit $n \rightarrow \infty$, 
\begin{eqnarray}
\lim_{n\rightarrow \infty} \hat{t}_i\hat{t}_i = \lim_{n\rightarrow \infty} \frac{1}{2}\left(\frac{1}{2} + \frac{1}{n}\right) = \frac{1}{4}.
\end{eqnarray}
Here the ``radius" does not depend on $n$ and we recover the standard $S^2$.

In $\mathcal{G}_n$, the operators $\hat{s}_i$ create and destroy same number of quanta and hence $\hat{s}_i: \mathcal{G}_n \rightarrow \mathcal{G}_n$. 
These are the generators of rotations in $\mathcal{G}_n$.
Using (\ref{irreducible_decomp}), we can decompose $\hat{s}_i$ as follows:
\begin{eqnarray}\label{sch_red_decomp_2}
\hat{s}_i = \hat{s}_i^{ee} \Lambda_{ee} + \hat{s}_i^{eo} \Lambda_{eo}+\hat{s}_i^{oe} \Lambda_{oe}+\hat{s}_i^{oo} \Lambda_{oo}.
\end{eqnarray}
Each $\hat{s}_i^\lambda$ satisfy
\begin{eqnarray}
\left[\hat{s}^{\lambda}_i, \hat{s}^{\lambda}_j\right]=i\epsilon_{ijk}  \hat{s}_k^{\lambda}       
\end{eqnarray}
and we obtain the following table:
\begin{eqnarray}\resizebox{4.5cm}{!}{$
\begin{array}{|c|c|}\hline &\\
\mathcal{F}_{2n}^{ee}& \hat{s}_i^{ee}\hat{s}_i^{ee} = \frac{n}{2}\left(\frac{n}{2}+1\right)\\  
&\\\hline&\\
\mathcal{F}_{2n+1}^{eo}& \hat{s}_i^{eo}\hat{s}_i^{eo} = \frac{n}{2}\left(\frac{n}{2}+1\right)\\  
&\\ \hline&\\
\mathcal{F}_{2n+1}^{oe}& \hat{s}_i^{oe}\hat{s}_i^{oe} = \frac{n}{2}\left(\frac{n}{2}+1\right)\\  
&\\ \hline&\\
\mathcal{F}_{2n+2}^{oo}& \hat{s}_i^{oo}\hat{s}_i^{oo} = \frac{n}{2}\left(\frac{n}{2}+1\right). \\  
&\\\hline
\end{array}$}
\end{eqnarray}

The subspaces $\mathcal{F}_{2n}^{ee}$, $\mathcal{F}_{2n+1}^{eo}$, $\mathcal{F}_{2n+1}^{oe}$ and $\mathcal{F}_{2n+2}^{oo }$ are all $(n+1)$-dimensional,  and each of them is 
the carrier space of $(n+1)$-dimensional UIR  of $SU(2)$. 

 On the other hand $\mathcal{G}_n$ is a $(4n+4)$-dimensional space and is  the carrier space of the reducible representation of $SU(2)$ generated by $\hat{s}_i$ with the irreducible decomposition (\ref{sch_red_decomp_2}). In this  space, $\hat{s}_i$ can be represented by $(4n+4) \times (4n+4)$ block diagonal matrices:

\begin{equation}\label{bloc_diag_s_i}\resizebox{8cm}{!}{$
 \hat{s}_i=\left(\hspace*{0.1cm}
  \begin{array}{ccccc}
\cline{1-1}
\bbord{  \textrm{Block}_{n+1}} & \abord   &     \hspace*{1cm}\bigzero  &         \\
\cline{1-1}
\cline{2-2}
&\bbord { \textrm{Block}_{n+1}} & \abord   &  &    \\
\cline{2-2}
\cline{3-3}
&&\bbord{ \textrm{Block}_{n+1}}  & \abord   &    \\
\cline{3-3}
\cline{4-4}
&\hspace*{-1cm}\bigzero &&\bbord{ \textrm{Block}_{n+1}}  & \abord \\
\cline{4-4}
  \end{array}\hspace*{-0.25cm}\right).$}
\end{equation}

\section{ Fuzzy line bundle with GBO}\label{section_4}

$\mathcal{H}_{nl}$ is the space of linear  maps $ \Phi: \mathcal{G}_n \rightarrow \mathcal{G}_l$, 
representated by $(4l+4)\times (4n+4)$ matrices.
It is a left $\mathcal{A}_l$-module and a right $\mathcal{A}_n$-module ($\mathcal{A}_l$ and $\mathcal{A}_n$ were defined in section \ref{section_1}).

On  $\mathcal{H}_{nn}$, the adjoint action of $s_i^{(n)}$ ($\hat{s}_i^{(n)}$ are the restriction of the operators $\hat{s}_i$ to $\mathcal{G}_n$) generates  rotations:
\begin{eqnarray}
Ad(\hat{s}_i) \Phi\equiv \hat{S}_i \Phi = \left[\hat{s}_i^{(n)}, \Phi\right], \quad \quad \Phi\in \mathcal{H}_{nn}.
\end{eqnarray}
In $\mathcal{H}_{nn}$, $\hat{S}_i$ defined above satisfies 
\begin{eqnarray}\label{lie_adj}
\left[\hat{S}_i, \hat{S}_j\right]=i \epsilon_{ijk}\hat{S}_k.
\end{eqnarray}

In $\mathcal{H}_{nl}$, the generators of the $SU(2)$  act as 
\begin{equation}
\hat{S}_i \Phi \equiv \hat{s}_i^{(l)} \Phi-\Phi \hat{s}_i^{(n)}, \quad \quad \Phi  \in \mathcal{H}_{nl}
\end{equation}
and  satisfy (\ref{lie_adj}).

The $\hat{S}_i$ acting  on $\mathcal{H}_{nl}$  correspond to the following reducible representation of  $SU(2)$ :
\begin{equation}\label{representation _reducible_adjoint}
\left(\frac{l}{2}\oplus \frac{l}{2} \oplus \frac{l}{2}\oplus\frac{l}{2}\right) \otimes \left(\frac{n}{2}\oplus \frac{n}{2} \oplus\frac{n}{2}\oplus \frac{n}{2}\right),
\end{equation}
which in turn can be decomposed into UIRs as 
\begin{eqnarray}\label{representation _reducible_adjoint_decomp}
\left[k\oplus (k+1)\ldots \oplus J\right] \oplus \left[k\oplus (k+1)\ldots \oplus J\right]\ldots_{16 \textrm{ copies}}, \quad \quad 
\end{eqnarray}
with $k=\frac{|l-n|}{2}$ and $  J=\frac{l+n}{2}$.

These sixteen identical copies of $\frac{l}{2}\otimes \frac{n}{2}$ essentially correspond to the following sixteen types of maps $\Phi: \mathcal{G}_n  \longrightarrow \mathcal{G}_l$ :
\begin{center}
\vspace{-.75cm}
\begin{equation}
\begin{tabular}{|c|c|}
\hline
$
\xymatrix@R=2pt{
 \\
      \mathcal{F}_{2n} \ar[r]\ar[dr]\ar[ddr]\ar[dddr]&    \mathcal{F}_{2l}\\
\mathcal{F}_{2n+1}&    \mathcal{F}_{2l+1}\\ 
\mathcal{F}_{2n+1}&    \mathcal{F}_{2l+1}\\
\mathcal{F}_{2n+2}&    \mathcal{F}_{2l+2}\\} 
$
 & 
$
\xymatrix@R=2pt{
 \\
      \mathcal{F}_{2n} &    \mathcal{F}_{2l}\\
\mathcal{F}_{2n+1}\ar[r]\ar[ur]\ar[dr]\ar[ddr]&    \mathcal{F}_{2l+1}\\ 
\mathcal{F}_{2n+1}&    \mathcal{F}_{2l+1}\\
\mathcal{F}_{2n+2}&    \mathcal{F}_{2l+2}\\} $
 \\&\\ \hline
$
\xymatrix@R=2pt{
 \\
      \mathcal{F}_{2n} &    \mathcal{F}_{2l}\\
\mathcal{F}_{2n+1}&    \mathcal{F}_{2l+1}\\ 
\mathcal{F}_{2n+1}\ar[r]\ar[uur]\ar[ur]\ar[dr]&    \mathcal{F}_{2l+1}\\
\mathcal{F}_{2n+2}&    \mathcal{F}_{2l+2}} $
&
$
\xymatrix@R=2pt{
 \\
      \mathcal{F}_{2n} &    \mathcal{F}_{2l}\\
\mathcal{F}_{2n+1}&    \mathcal{F}_{2l+1}\\ 
\mathcal{F}_{2n+1}&    \mathcal{F}_{2l+1}\\
\mathcal{F}_{2n+2}\ar[r]\ar[uuur]\ar[ur]\ar[uur]&    \mathcal{F}_{2l+2}\\} $ \\  &\\ \hline
\end{tabular}\label{map_table_1}
\end{equation}
\end{center}
 \vspace*{0cm}

The maps  $\Phi \in \mathcal{H}_{nl}$ can be expanded in the basis of the eigenfunctions of $\hat{S}_i\hat{S}_i$ and $\hat{S}_3$ belonging to the irreducible representations in decomposition of (\ref{representation _reducible_adjoint_decomp}).
We can explicitly find the basis vectors by constructing the highest weight vectors, and by repeated action of   $ \hat{S}_- (\equiv \hat{S}_1-i \hat{S}_2)$.
Below we list all the highest weight vector (details for a specific example are in the appendix \ref{app_red}) :

\begin{center}
\vspace*{-2cm}
  \resizebox{11cm}{!}{ 
\begin{tabular}[scale=.2]{|c|l|c|c|c|}\hline
&\\
 & Highest  weight function     \\ 
&$$\\\hline &\\
$\mathcal{F}_{2n}^{ee}\rightarrow\mathcal{F}_{2l}^{ee}$ & $\left(\Psi^{ee\rightarrow ee}\right)^{Jk}_{j,j}=\left(\hat{b}_{1}^{e\dagger}\right)^{\tilde{l}}\left(\hat{b}_{2}^{e}\right)^{\tilde{n}}\Lambda_{ee}$  \\[1ex]\hline&\\
 $\mathcal{F}_{2n}^{ee}\rightarrow\mathcal{F}_{2l+1}^{eo}$&$\left(\Psi^{ee\rightarrow eo}\right)^{Jk}_{j,j}=\hat{a}_2^\dagger \frac{1}{\sqrt{\hat{N}_2+1}} \left(\hat{b}_{1}^{e\dagger}\right)^{\tilde{l}} \left(\hat{b}_{2}^{e}\right)^{\tilde{n}}\Lambda_{ee}$ \\ [1ex] \hline &\\
$\mathcal{F}_{2n}^{ee}\rightarrow\mathcal{F}_{2l+1}^{oe}$ & $\left(\Psi^{ee\rightarrow oe}\right)^{Jk}_{j,j}=\hat{a}_1^\dagger \frac{1}{\sqrt{\hat{N}_1+1}} \left(\hat{b}_{1}^{ e\dagger}\right)^{\tilde{l}} \left(\hat{b}_{2}^{e}\right)^{\tilde{n}}\Lambda_{ee}$  \\ [1ex] \hline &\\
 $\mathcal{F}_{2n}^{ee}\rightarrow\mathcal{F}_{2l+2}^{oo}$&  $\left(\Psi^{ee\rightarrow oo}\right)^{Jk}_{j,j}=\hat{a}_1^\dagger \frac{1}{\sqrt{\hat{N}_1+1}} \hat{a}_2^\dagger \frac{1}{\sqrt{\hat{N}_2+1}} \left(\hat{b}_{1}^{ e\dagger}\right)^{\tilde{l}} \left(\hat{b}_{2}^{e}\right)^{\tilde{n}}\Lambda_{ee}$  \\[1ex] \hline&\\
$\mathcal{F}_{2n+1}^{eo}\rightarrow\mathcal{F}_{2l}^{ee}$ &  $\left(\Psi^{eo\rightarrow ee}\right)^{Jk}_{j,j}=\hat{a}_2 \frac{1}{\sqrt{\hat{N}_2}} \left(\hat{b}_{1}^{ e\dagger}\right)^{\tilde{l}} \left(\hat{b}_{2}^{o}\right)^{\tilde{n}}\Lambda_{eo}$ \\ [1ex]\hline &\\
$\mathcal{F}_{2n+1}^{eo}\rightarrow\mathcal{F}_{2l+1}^{eo}$ & $\left(\Psi^{eo\rightarrow eo}\right)^{Jk}_{j,j}=\left(\hat{b}_{1}^{e\dagger}\right)^{\tilde{l}}\left(\hat{b}_{2}^{o}\right)^{\tilde{n}}\Lambda_{eo}$   \\[1ex] \hline &\\
$\mathcal{F}_{2n+1}^{eo}\rightarrow\mathcal{F}_{2l+1}^{oe}$ &$\left(\Psi^{eo\rightarrow oe}\right)^{Jk}_{j,j}=\hat{a}_1^\dagger \frac{1}{\sqrt{\hat{N}_1+1}}\hat{a}_2 \frac{1}{\sqrt{\hat{N}_2}} \left(\hat{b}_{1}^{ e\dagger}\right)^{\tilde{l}} \left(\hat{b}_{2}^{o}\right)^{\tilde{n}}\Lambda_{eo}$  \\ [1ex]\hline&\\
$\mathcal{F}_{2n+1}^{eo}\rightarrow\mathcal{F}_{2l+2}^{oo}$ &   $\left(\Psi^{eo\rightarrow oo}\right)^{Jk}_{j,j}=\hat{a}_1^\dagger \frac{1}{\sqrt{\hat{N}_1+1}} \left(\hat{b}_{1}^{ e\dagger}\right)^{\tilde{l}} \left(\hat{b}_{2}^{o}\right)^{\tilde{n}}\Lambda_{eo}$    \\[1ex] \hline &\\
 $\mathcal{F}_{2n+1}^{oe}\rightarrow\mathcal{F}_{2l}^{ee}$&    $\left(\Psi^{oe\rightarrow ee}\right)^{Jk}_{j,j}=\hat{a}_1\frac{1}{\sqrt{\hat{N}_1}} \left(\hat{b}_{1} ^{o\dagger}\right)^{\tilde{l}} \left(\hat{b}_{2}^{e}\right)^{\tilde{n}}\Lambda_{oe}$  \\ [1ex]\hline &\\
 $\mathcal{F}_{2n+1}^{oe}\rightarrow\mathcal{F}_{2l+1}^{eo}$ & $\left(\Psi^{oe\rightarrow eo}\right)^{Jk}_{j,j}=\hat{a}_1\frac{1}{\sqrt{\hat{N}_1}}\hat{a}_2^\dagger \frac{1}{\sqrt{\hat{N}_2+1}} \left(\hat{b}_{1} ^{o\dagger}\right)^{\tilde{l}} \left(\hat{b}_{2}^{e}\right)^{\tilde{n}}\Lambda_{oe}$    \\[1ex] \hline&\\
  $\mathcal{F}_{2n+1}^{oe}\rightarrow\mathcal{F}_{2l+1}^{oe}$& $\left(\Psi^{oe\rightarrow oe}\right)^{Jk}_{j,j}=\left(\hat{b}_{1}^{o\dagger}\right)^{\tilde{l}}\left(\hat{b}_{2}^{e}\right)^{\tilde{n}}\Lambda_{oe}$  $\frac{(\tilde{l}+\tilde{n})}{2}$\\ [1ex] \hline &\\
 $\mathcal{F}_{2n+1}^{oe}\rightarrow\mathcal{F}_{2l+2}^{oo}$ & $\left(\Psi^{oe\rightarrow oo}\right)^{Jk}_{j,j}=\hat{a}_2^\dagger \frac{1}{\sqrt{\hat{N}_2+1}} \left(\hat{b}_{1} ^{o\dagger}\right)^{\tilde{l}} \left(\hat{b}_{2}^{e}\right)^{\tilde{n}}\Lambda_{oe}$    \\ [1ex]\hline &\\
 $\mathcal{F}_{2n+2}^{oo}\rightarrow\mathcal{F}_{2l}^{ee}$ &  $\left(\Psi^{oo\rightarrow ee}\right)^{Jk}_{j,j}=\hat{a}_1\frac{1}{\sqrt{\hat{N}_1}} \hat{a}_2 \frac{1}{\sqrt{N}_2}\left(\hat{b}_1^{o\dagger}\right)^{\tilde{l}} \left(\hat{b}_2^{o}\right)^{\tilde{n}}$   \\[1ex] \hline &\\
$\mathcal{F}_{2n+2}^{oo}\rightarrow\mathcal{F}_{2l+1}^{eo}$ &$\left(\Psi^{oo\rightarrow eo}\right)^{Jk}_{j,j}=\hat{a}_1\frac{1}{\sqrt{\hat{N}_1}} \left(\hat{b}_1^{o\dagger}\right)^{\tilde{l}} \left(\hat{b}_2^{o}\right)^{\tilde{n}}$     \\[1ex] \hline &\\
$\mathcal{F}_{2n+2}^{oo}\rightarrow\mathcal{F}_{2l+1}^{oe}$ &  $\left(\Psi^{oo\rightarrow oe}\right)^{Jk}_{j,j}= \hat{a}_2 \frac{1}{\sqrt{N}_2}\left(\hat{b}_1^{o\dagger}\right)^{\tilde{l}} \left(\hat{b}_2^{o}\right)^{\tilde{n}}$  \\[1ex] \hline &\\
$\mathcal{F}_{2n+2}^{oo}\rightarrow\mathcal{F}_{2l+2}^{oo}$ & $\left(\Psi^{oo\rightarrow oo}\right)^{Jk}_{j,j}=\left(\hat{b}_1^{o\dagger}\right)^{\tilde{l}} \left(\hat{b}_2^{o}\right)^{\tilde{n}}$\\ [1ex]\hline
\end{tabular}}
\end{center}~where $0\leq \tilde{n} \leq n$ and $\tilde{l}-\tilde{n}=l-n=2k$. 

Each of these highest weight vectors belong to the representation with 
\begin{equation}
j=\frac{1}{2}(\tilde{l}+\tilde{n}).
\end{equation}
The range of $j$ in all the cases is
\begin{equation}
j=k,k+1,k+2\ldots J.
\end{equation}

These are the irreducible representations  in (\ref{representation _reducible_adjoint_decomp}).
The other basis vectors can be found by operating  the lowering operator suitably, as in (\ref{lowering1}), on these highest weight vectors.

Therefore any arbitrary  element of $\mathcal{H}_{nl}$ can be expanded in the basis of these operators as
\begin{eqnarray}
\Phi &=& \sum_\alpha \sum_{j=k}^J \sum_{m=-j}^j  C^{j,m}_{\alpha} \left(\Psi^{\alpha}\right)^{Jk}_{j,m}
\end{eqnarray}
where $\alpha = ee\rightarrow ee, ee\rightarrow oo \cdots$.


The topological charge operator $\hat{K}_0$ is
\begin{equation}
\hat{K}_0 \equiv = \frac{1}{2} \left[ \hat{M},\quad \right], \quad \quad \hat{M}= \hat{b}_1^\dagger\hat{b}_1 + \hat{b}_2^\dagger \hat{b}_2.
\end{equation}
The elements of  $\mathcal{H}_{nl}$ satisfy
\begin{equation}
\hat{K}_0 \Phi = \frac{k}{2} \Phi, \quad\quad \Phi \in \mathcal{H}_{nl},\quad k \in \mathbb{Z}^+.
\end{equation}
Hence $\Phi \in  \mathcal{H}_{nl}$ are the noncommutative analogue of the complex line bundles with topological charge $k$.

\section{Mixed states}\label{section_6}

The algebra ${\cal A}$ of the fuzzy sphere is generated by $\{ \mathbb{I}, \hat{s}_i, \hat{s}_i \hat{s}_j, \cdots\}$ where the $\hat{s}_i$ were defined in (\ref{js_n}). 
This algebra is also a $*$-algebra, with $\hat{s}_i^* = \hat{s}_i$. In general, this algebra is simply the algebra $M_{N+1}(\mathbb{C})$ of $(N+1)\times (N+1)$ matrices. 
Notice that this includes the description of both the irreducible as well as the reducible fuzzy spheres.

A state on this algebra is a linear map $\omega: {\cal A} \rightarrow \mathbb{C}$ satisfying 
$$\omega(A^*A) \geq 0  \quad \forall A \in {\cal A}, \quad\quad \quad \omega(\mathbb{I}) =1.$$ 

Given a state on a $*$-algebra, we can use the GNS construction to construct the Hilbert space $\mathds{H}_{GNS}$ to make contact with 
standard quantum mechanical description. The advantage of this algebraic formalism is that both pure and mixed density matrices may be 
discussed in the same unified language.

To describe the fuzzy sphere in an algebraic language, we need to impose more conditions on the state $\omega$. To describe the irreducible $S_F^2$, we require that $\omega(\hat{s}_i \hat{s}_i) = \frac{N}{2}\left(\frac{N}{2}+1\right)\omega(\mathbb{I})=\frac{N}{2}\left(\frac{N}{2}+1\right)$. With this 
condition, the resulting $\mathds{ H}_{GNS}$ is simply the unique (upto unitary equivalence) carrier space of the $(N+1)$-dimensional representation 
of $SU(2)$.

To describe the reducible representations, consider $n$ projectors $P_a^{(n_a)}$, $a=1,\cdots n$, where $n_a$ is the rank of $P_a^{(n_a)}$. The projectors 
satisfy $P_a^{(n_a)} P_b^{(n_b)} = \delta_{ab} P_i^{(n_a)}$ and $\sum_{a=1}^n P_a^{(n_a)} = \mathbb{I}_{N+1}$. A state $\omega_{a}$ which 
satisfies 
\begin{equation}
\omega_{a}(P_a^{(n_a)} \hat{s}_i\hat{s}_i P_a^{(n_a)}) = s^{(a)} (s^{(a)}+1)
\end{equation}
gives us the $\mathds{ H}_{GNS}$ corresponding to the reducible representation $s^{(1)} \oplus s^{(2)} \cdots \oplus s^{(n)}$. The algebra $\mathcal{A}$ splits into 
$\sum_a \mathcal{A}^{(a)}$.

If all the $s^{(a)}$ are distinct, then the projectors $P_a^{(n_a)}$ are unique (upto unitary equivalence). This is not so when some of the $s^{(a)} (> 0)$ 
are repeated. Let us illustrate this with an example. Consider the situation when $s^{(1)}=s^{(2)} =s>0$. Let us label the states by 
$|s, s_3; \alpha), |s_3|\leq s, \alpha=1,2$. The projector $P_\alpha$ is given by
\begin{equation}
P_\alpha = \sum_{s_3=-s}^s |s, s_3; \alpha)(s, s_3; \alpha|.
\end{equation}
These projectors are not unique. Under the transformation 
\begin{equation}
|s, s_3; \alpha) \rightarrow  \sum_\alpha u_{\beta\alpha}|s, s_3; \alpha), \quad\quad u \in U(2),
\end{equation}
it is easy to check that
\begin{equation}
P_\alpha \rightarrow P[u;\alpha] = \sum_{s_3=-s}^s\sum_{\beta,\gamma}u^\dagger_{\gamma\alpha} u_{\alpha\beta}|s, s_3; \beta)(s, s_3; \gamma|
\end{equation}
is still a projector but it does not project to the fixed subspace $\{|s, s_3; \alpha)\}$.

It is important to recognise that these projectors are not elements of the observable algebra: there is no canonical construction of $P_\alpha$ using 
only the elements of the algebra $\mathcal{A}$.

The algebra  (\ref{nc_fuzzy}) is a $(4n+4)$-dimensional reducible representation of $SU(2)$: it contains four identical copies 
of the $(n+1)$-dimensional irreducible representation. The carrier space of the reducible representation can be splited as  (\ref{split_g_n}).
The states in $\mathcal{G}_n$ are labelled as
\begin{equation}
|s, s_3 ; \alpha ), 
\end{equation}
where $\alpha= ee,eo,oe,oo$ labels the subspaces $F_{2n}^{ee}, F_{2n+1}^{eo}, F_{2n+1}^{oe}$ and $ F_{2n+2}^{oo}$ respectively.
For example in $\mathcal{G}_1$,  $s=\frac{1}{2}$ and the states
\begin{equation}
|2,0\rangle = |s=\frac{1}{2}, s_3=\frac{1}{2} ; ee ), \quad\quad |3,0\rangle = |s=\frac{1}{2}, s_3=\frac{1}{2} ; oe )
\end{equation}
belong to the irreducible subspaces $\mathcal{F}_2^{ee}$ and  $\mathcal{F}_3^{oe}$ respectively.

The projectors $\Lambda^{\alpha}$ projects to these irreps. To indentify the specific irreps that a state belongs to,  we need to know 
\begin{equation}
(s,s_3;\alpha|\Lambda^\beta|s, s_3; \alpha)  = \delta_{\alpha\beta}
\end{equation}
But these projectors are not unique. $\Lambda^\alpha$ are given in terms of the  $\hat{N}_1$ $(=\hat{a}_1^\dagger\hat{a}_1)$ and $\hat{N}_2$ $(=\hat{a}_2^\dagger\hat{a}_2)$.
But $\hat{N}_1$  and $\hat{N}_2$ , and hence the projectors $\Lambda^\alpha$, are not elements of the algebra of observables (\ref{nc_fuzzy}).

As all the irreps are identical,  the expectation value of any element $A$ of the algebra of observables in all the irreps are the same:
\begin{equation}
\omega_\alpha(A) = \sum_{s_3} \sum_{s_3'} (s,s_3;\alpha |A| s, s_3';\alpha), \quad \quad \forall \alpha.
\end{equation}
Therefore with a probability vector $\lambda_\alpha$ ($0\leq\lambda_\alpha\leq 1$, \,\,$\sum_\alpha \lambda_\alpha =1$), we can define 
\begin{equation}
\omega( A) = \sum_\alpha \lambda_\alpha\omega_\alpha(A)  = \omega_\alpha(A), \quad\quad \forall \alpha.
\end{equation}

For any  transformation 
\begin{equation}\label{gauge_trans1}
|s, s_3: \beta)= \sum_\alpha u_{\beta\alpha}|s, s_3: \alpha),
\end{equation}
$u$ belongs to $U(4)$ because the states are orthonormal and the  spaces of the UIRs are invariant subspaces.

Further, under the transformation (\ref{gauge_trans1}), 
\begin{equation}
\omega(A) \rightarrow \omega(A).
\end{equation}
The state $\omega(A)$ remains invariant, and the system has $U(4)$ gauge symmetry.

Under the transformation  (\ref{gauge_trans1}), 
\begin{equation}
\lambda_\beta (u) = \sum_{\alpha} \lambda_\alpha |u_{\alpha \beta}|^2
\end{equation}


The state $\omega_\alpha(A)$  can be written as
\begin{equation}
\omega_\alpha(A) = Tr \left[\rho^\alpha A\right], \quad \quad \rho^\alpha \equiv \sum_{s_3} p_{s_3} |s, s_3: \alpha)(s,s_3:\alpha|.
\end{equation}
Similarly, $\omega(A)$  can be written as
\begin{equation}
\omega(A)=Tr \left[\rho A\right].
\end{equation}
The two density matrices are related as
\begin{equation}\label{density_mixed}
\rho\equiv \sum_\alpha \lambda_\alpha\rho^\alpha, \quad \quad 0<\lambda_\alpha< 1, \quad  \sum_\alpha\lambda_\alpha=1.
\end{equation}
The  decomposition (\ref{density_mixed}) is not unique. This non-uniqueness is parametrized by the unitary matrices $u \in U(4)$.
Therefore the state of the fuzzy sphere (\ref{nc_fuzzy}) is mixed which is invariant under the gauge group $U(4)$.

The fuzzy sphere (\ref{nc_fuzzy}) being in a mixed state have dramatic consequences.  In \cite{Myers:1999ps}, it was shown that the fuzzy sphere in a reducible representation has  higher energy than that in the irreducible representation, and that the reducible representations are metastable vacua.
Our results show that this discussion needs to be refined: when the fuzzy sphere is in a reducible representation (with several identical copies of some particular irrep), then the corresponding quantum state is necessarily mixed. On the other hand the fuzzy sphere in the irreducible representation is in a pure state.  The mixed states can never evolve to a pure state under a unitary time evolution, and a decay to the minimum energy configuration is not possible. Such a decay is possible only if one enlarges the algebra of observables.

Though  $Tr[\rho_\alpha A]$  is same in all the irreps, the von Neumann entropy is not:
\begin{equation}
-Tr[\rho_\alpha \log \rho_\alpha] \neq -Tr[\rho_\beta \log \rho_\beta], \quad \quad \alpha\neq \beta.
\end{equation}
This ambiguity in the definition leads to an intrinsic non-zero entropy in the mixed state.
For the density matrix defined in (\ref{density_mixed}),  the von Neumann entropy is given by
\begin{equation}
S= - \sum_\alpha \lambda_\alpha (u) \log \lambda_\alpha (u)
\end{equation}

The map $\lambda_\alpha \rightarrow \lambda_\alpha (u)$ is a Markovian:
\begin{eqnarray}
\lambda_\beta(u)= \sum_\alpha \lambda_\alpha T_{\alpha\beta},
\end{eqnarray}
and the matrix $T$ satisfying
\begin{eqnarray}
T_{\alpha\beta}=  |u_{\alpha\beta}|^2 \geq 0, \quad  \sum_\alpha T_{\alpha\beta} =1, \quad  \sum_\beta T_{\alpha\beta} =1,
\end{eqnarray}
is a doubly stochastic matrix. Being stochastic, it guarantees that the Markov process is irreversible and that the entropy of the system is driven to its 
 maximum value. 
The entropy is maximized when $\lambda_{ee}=\lambda_{eo}=\lambda_{oe}=\lambda_{oo}= \frac{1}{4}$ and 
\begin{eqnarray}
S_{max}=2 \log 2.
\end{eqnarray}
The above formula for $S_{max}$ is for the specific situation when the $S_F^2$ algebra has four identical irreps. The more general formula is derived later.

\subsection{Monopole states are mixed}


We showed in section \ref{section_4} that fuzzy monopoles are described by the $SU(2)$-representation
\begin{eqnarray}\label{representation _reducible_adjoint_decomp1}
\left[k\oplus (k+1)\cdots \oplus J\right] \oplus \left[k\oplus (k+1)\ldots \oplus J\right]\cdots \;\;_{16\,\, \textrm{copies}}.
\end{eqnarray}
The states in distinct irreps above belong to different super-selection sectors. But the states belonging to the identical representations (say, $j=k$ in copy-1 and $j=k$ in copy-2) cannot be distinguished by any observable of the algebra, because the projectors are not elements of this algebra.
These states are 16-fold degenerate with a $U(16)$ gauge symmetry, and are mixed. As before, we can write a density matrix 
\begin{equation}
\tilde{\rho} = \sum_{\mu} \tilde{\lambda}_\mu (\tilde{u}) \tilde{\rho}^\mu, \quad\quad u\in U(16), \quad \quad \mu=ee\rightarrow ee, ee\rightarrow eo\cdots
\end{equation}

The entropy of this mixed state is
\begin{equation}
\tilde{S}= \sum_\mu -\tilde{\lambda}_\mu \log \tilde{\lambda}_\mu.
\end{equation}

The doubly stochastic process drives the system to a configuration with maximum entropy
\begin{equation}
\tilde{S}_{max} = 4 \log 2.
\end{equation}


More generally, we can consider the oscillators (\ref{b_red})
\begin{equation}
 \hat{b}_1^{(K_1)} = \sum_{i=0}^{K_1-1} \hat{b}_1^i \Lambda_1^i,  \quad\quad  \hat{b}_2^{(K_2)} = \sum_{i=0}^{K_2-1} \hat{b}_2^i \Lambda_2^i
\end{equation}
where $K_1$ and $K_2$ are natural numbers.

When $K_1K_2 >1$, the representations of the $SU(2)$ algebra (\ref{nc_fuzzy}) are in general reducible and contain identical $K_1K_2$ irreps. Each state is $K_1K_2$-fold degenerate. The projectors $\Lambda_1^i$ and $\Lambda^i_2$ are not elements of the observable algebra and hence cannot be used distinguish between the irreps.
As a result, the fuzzy sphere is in a mixed state 
and the maximum value of its entropy is 
\begin{equation}
S_{max} = \log (K_1 K_2).
\end{equation}

A similar analysis holds for the fuzzy monopoles as well. 
The monopole bundle corresponds to a quantum mixed state, with von Neumann  entropy
\begin{equation}
\tilde{S} = \sum_{\mu= 1}^{(K_1K_2)^2}  - \tilde{\lambda}_\mu \log \tilde{\lambda}_\mu.
\end{equation}
The entropy is maximized when $\tilde{\lambda}_\mu = \frac{1}{(K_1 K_2)^2}$:
\begin{equation}
\tilde{S}_{max} = \log (K_1 K_2)^2 = 2\log (K_1 K_2).
\end{equation}

\appendix
\section{Appendix}\label{app_red}
Let $\mathcal{H}_{nl}^{ee \rightarrow ee}$  be the space of maps $\Phi^{ee\rightarrow ee}: \mathcal{F}^{ee}_{2n} \rightarrow \mathcal{F}^{ee}_{2l}$. $\mathcal{H}_{nl}^{ee \rightarrow ee}$ is also a bimodule which is a subset of $\mathcal{H}_{nl}$. In this subspace, the generators of $SU(2)$ act as
\begin{equation}
\hat{S}_i \Phi \equiv \hat{s}_i^{(l^{ee})} \Phi^{ee\rightarrow ee}-\Phi^{ee\rightarrow ee} \hat{s}_i^{(n^{ee})}, \quad \quad \Phi^{ee\rightarrow ee} \in \mathcal{H}^{ee\rightarrow ee}_{nl}.
\end{equation}
 where $ \hat{s}_i^{(n^{ee})}$ is the restriction of $\hat{s}_i$ in $\mathcal{F}^{ee}_{2n} $:
\begin{eqnarray}
  \hat{s}_+^{ee}= \hat{b}^{e\dagger}_{1} \hat{b}^e_{2},\quad\quad\hat{s}_-^{ee}= \hat{b}^{e\dagger}_{2} \hat{b}^e_{1}, \quad\quad \hat{s}_3^{ee}= \frac{1}{2}\left(\hat{b}^{e\dagger}_{1} \hat{b}^e_{1}- \hat{b}^{e\dagger}_{1} \hat{b}^e_{2}\right).
\end{eqnarray}

Let us consider the operator $\hat{f}_1=\left(\hat{b}_{1+}^{\dagger}\right)^{\tilde{l}}\left(\hat{b}_{2+}\right)^{\tilde{n}}\Lambda_{ee}$:
\begin{eqnarray}
&&\hat{f}_1: \mathcal{F}_{2n}^{ee} \rightarrow \mathcal{F}_{2l}^{ee} \quad \textrm{if }\quad  \tilde{l}-\tilde{n}=l-n,  \quad 0\leq\tilde{n}\leq n \\
&\textrm{with}& k=\frac{l-n}{2} \quad\textrm{  and  } \quad J=\frac{l+n}{2}  .
\end{eqnarray}

It is easy to check that 
\begin{eqnarray}
&&\hat{S}_+ \hat{f} \equiv \left[\hat{s}_{+}^{ee}, \hat{f}_1\right] =0,\\
&&\hat{S}_3 \hat{f} \equiv \left[\hat{s}_3^{ee}, \hat{f}_1\right]=\frac{1}{2}\left(\tilde{l}+\tilde{n}\right)\hat{f}_1.
\end{eqnarray}
So $\hat{f}_1$ is the highest weight function belonging to the spin-$j$ representation:
\begin{eqnarray}
j=\frac{1}{2}\left(\tilde{l}+\tilde{n}\right).
\end{eqnarray}
Let us denote this highest weight function by $\left(\Psi^{ee\rightarrow ee}\right)^{Jk}_{jj}$:
\begin{eqnarray}
\left(\Psi^{ee\rightarrow ee}\right)^{Jk}_{jj}=\left(\hat{b}_{1}^{e\dagger}\right)^{\tilde{l}}\left(\hat{b}_{2}^e\right)^{\tilde{n}}\Lambda_{ee}.
\end{eqnarray}
The lower weight functions $\left(\Psi^{ee\rightarrow ee}\right)^{Jk}_{jm}$ can be obatined by repeated action of the lowering operator:
\begin{eqnarray} \label{lowering1}
&&\left(\hat{S}_-\right)^{j-m} \left(\Psi^{ee\rightarrow ee}\right)^{Jk}_{jj} = c_{jm}^{Jk} \left(\Psi^{ee\rightarrow ee}\right)^{Jk}_{jm}, \quad \quad  c_{jm}^{Jk}\in \mathbb{C}
\end{eqnarray}
where $\hat{S}_- \equiv  \left[\hat{s}_{+}^{ee}, \,\,\, \right].$

The range of $j$ can easily be computed but putting the suitable values of $\tilde{n}$ and $\tilde{l}$:
\begin{equation}
j=k, k+1, k+2 \ldots J.
\end{equation}

\end{document}